\documentstyle[sprocl]{article}

\input{psfig.sty}

\def\Journal#1#2#3#4{{#1} {\bf #2}, #3 (#4)}

\def\NPB{{\em Nucl. Phys.} B}

\def\PRD{{\em Phys. Rev.} D}
\def\PRB{{\em Phys. Rev.} B}

\def\be{\begin{equation}}
\def\ee{\end{equation}}
\def\ba{\begin{array}}
\def\ea{\end{array}}
\def\bea{\begin{eqnarray}}
\def\eea{\end{eqnarray}}

\begin{document}

\title{FLOW EQUATIONS AND THE CHIRAL PHASE TRANSITION}

\author{B.-J. SCHAEFER, O. BOHR AND J. WAMBACH}

\address{Institut f\"ur Kernphysik, TU Darmstadt,
Schlossgartenstr. 9\\  D-64289 Darmstadt\\  Germany}

\maketitle\abstracts{Flow equations for an $O(N)$--symmetric effective potential
are discussed and solved for the finite temperature case. The model is
investigated at the critical point and critical exponents for various
$N$ are calculated.}

\section{Introduction}
The $N$-component scalar field theory with an $O(N)$-symmetry often
serves as a prototype for symmetry restoration investigations at
finite temperature. The manifestation of universality of critical
phenomena enables the applicability of the $O(N)$-symmetric model to a
wide class of very different physical systems around the critical
temperature.

Wilczek and Pisarski have argued that the two flavor chiral transition
is either of second order or of first order depending on the $U_A (1)$
anomaly. In case that the transition is of second order Wilczek and
Rajagopal have shown that two flavor QCD belongs to the same
universality class as the three dimensional four component Heisenberg
magnet~\cite{wilc}. Thus for $N=4$ this theory is used as an effective
model for the chiral phase transition in two flavor QCD enabling
deeper insights into the chiral phase transition in QCD provided that
the chiral order parameter remains close to thermal equilibrium
through the transition. On the lattice it is still not clear whether
QCD for two flavors and the $O(4)$-model lie in the same universality
class~\cite{laer}.

In a second order transition, the system is at an infrared fixed point
of the renormalization group. This means that the physics is scale
invariant and the flow equations show a scaling solution. Due to a
diverging correlation length in the chiral limit the order parameter
fluctuates on all length scales and the correlation functions exhibit
a power law behavior with critical exponents.

The article is organized as follows: In Sect.~\ref{floweq} we
summarize the concept of the renormalization group (RG) method. In
Sect.~\ref{finiteT} we present some results for the $O(4)$-model at
finite temperature.  Finally Sect.~\ref{criticalbeh} is dedicated to
the critical behavior of the system at exactly the critical
temperature.

\section{Concept of the Renormalization Group Method}
\label{floweq}

RG flow equations predict the behavior of a given theory in different
momentum regimes. The main ingredient of the RG approach is the
integration of irrelevant short-distance modes in order to derive a
low-energy effective theory in the infrared in analogy to a discrete
block-spin transformation on the lattice. For this purpose one
introduces an infrared scale $k$ that separates the fast-fluctuating
short-distance modes from the slowly-varying modes. This idea --
pioneered by Wilson and Kadanoff in the early seventies~\cite{kada} --
can be systematically incorporated e.g.~in a momentum regularization
in a transparent way. It results in an effective action parameterized
by the averaged blocked field at the scale $k$. Thus the full
one-particle irreducible (1PI) Feynman graphs or vertex functions are
generated by the renormalized effective action in the limit $k \to 0$
and the $k$-dependent effective action provides a smooth interpolation
between the bare action defined at the UV scale $\Lambda$, where no
fluctuations are considered, and the renormalized effective action in
the IR\@. The flow pattern of the given theory is obtained by studying
an infinitesimal change of the IR scale $k$ in the effective
$k$-dependent action and is governed by differential flow equations.

Of course, the integration step cannot be performed in an exact way
and one has to resort to some approximation such as the loop
expansion. However a sharp momentum cutoff is in conflict with
important underlying symmetries of the considered theory. The task is
then to implement both the UV and the IR cutoff scales in a
symmetry-conserving way, which can be accomplished by an operator
cutoff regularization in Schwinger's proper-time representation.

The one-loop contribution to the effective action in general yields a
non-local UV diverging logarithm, which can be rewritten in a
proper-time representation and leads to
\be
\label{eq}
\Gamma = -\frac{1}{2} \int d^4 x \int\limits_{1/\Lambda^2}^\infty
\frac{d\tau}{\tau}
\int\frac{d^4 q}{(2\pi)^4} \mbox{Tr} e^{-\tau (q^2 + V'')}
\ee
where the primes on the potential $V$ denote differentiation
w.r.t.~$\phi$. Here, the UV (IR) divergences appear for $\tau \to 0$
($\infty$).  We modify the above expression by introducing a
regulating $a$ $priori$ unknown smearing function $f_k (\tau)$ in
the proper-time integrand. Differentiating the resulting $k$-dependent
effective action w.r.t.\ the scale $k$ yields the flow equations. It
can be shown that all universal results in the infrared are
independent of the choice of this regulating smearing
function~\cite{papp}.

\section{Finite Temperature}
\label{finiteT}
It is straightforward to generalize the above approach to finite
temperature. Within the imaginary time (Matsubara) approach the
four-dimensional theory at $T=0$ is conveniently matched to the
effectively three-dimensional behavior at the critical temperature by
a replacement of the four-dimensional momentum integration in
Eq.~(\ref{eq}) with a three-dimensional integration and a Matsubara
summation over the corresponding bosonic frequencies. Thus at finite
temperature the scale $k$ serves as a generalized IR cutoff for a
combination of the three-dimensional momenta and Matsubara
frequencies.

To solve the flow equations numerically we start the evolution in the
broken phase deep in the UV region $k = \Lambda$ and use a tree-level
parameterization of the potential with two initial values. These
initial values are fixed at $T=0$ and are constant in a relatively
large temperature region.

\begin{figure}[htbp]
\vspace{-0.1in}
\hspace{2.5cm}
\centerline{\hbox{
\psfig{file=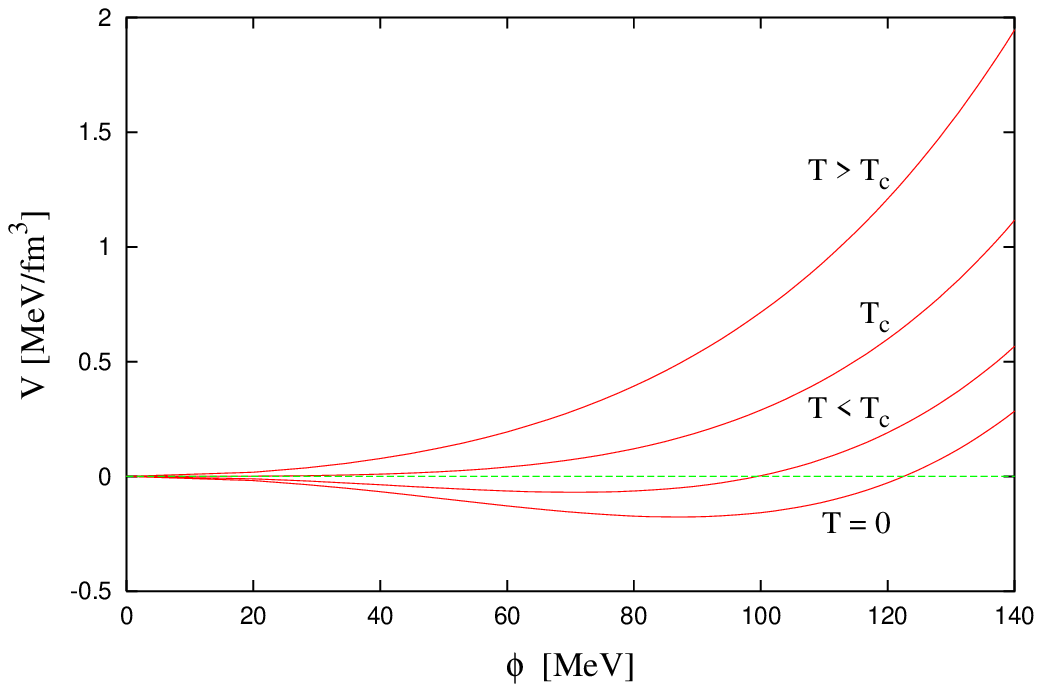,height=1.9in,width=2.45in}
\psfig{file=space.ps,width=0.05cm,angle=-90}
\psfig{file=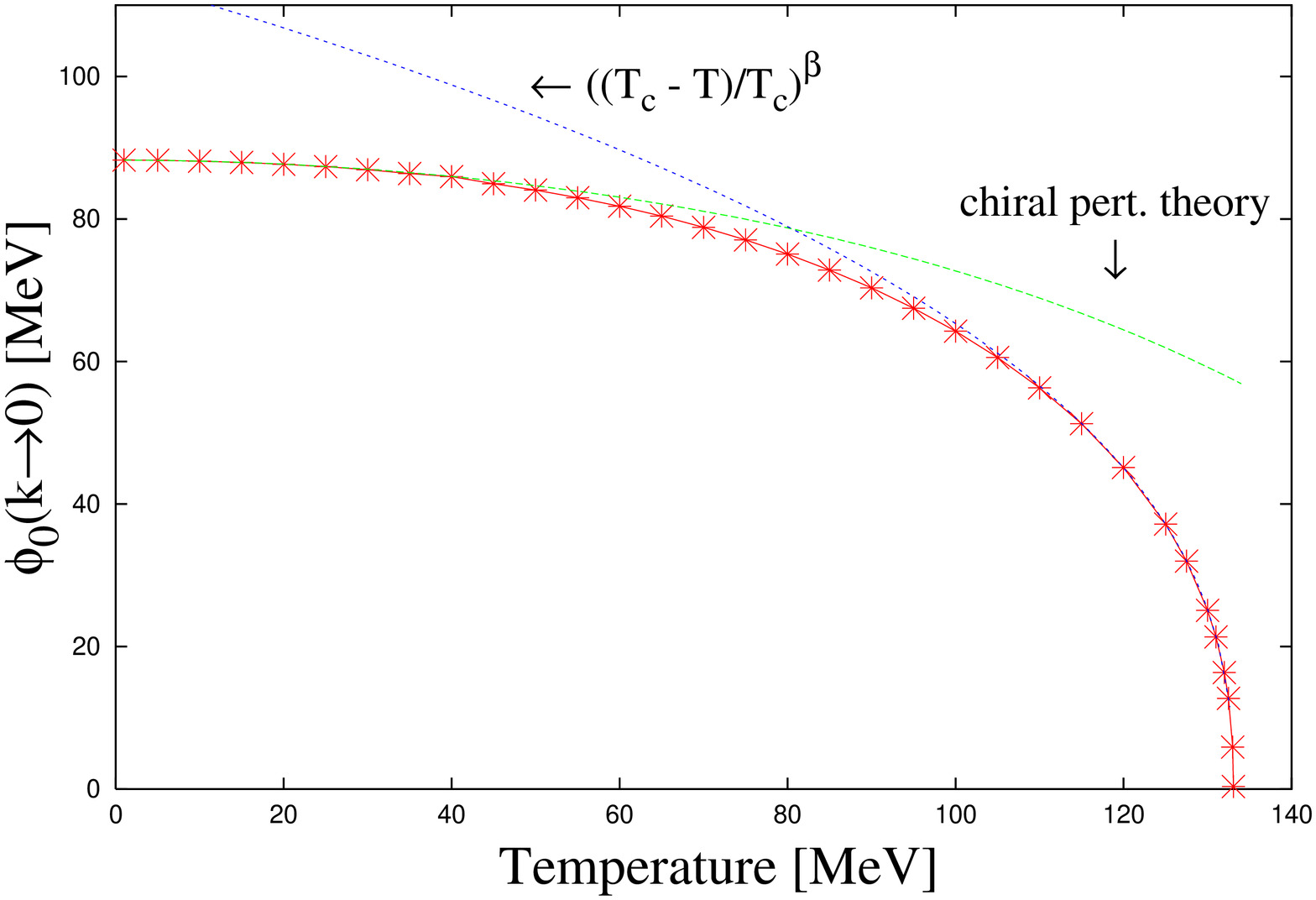,height=1.9in,width=2.4in,angle=-90}
}}
\parbox{12cm}
{\caption{\label{figcoupling} Left panel: The potential for various
temperatures. Right panel: The temperature-dependent order parameter in 
the chiral limit.}}
\end{figure}

In the $k\to 0$ limit the temperature dependence of the potential
$V(\phi )$ (left panel of Fig.~\ref{figcoupling}) and the order
parameter $\phi_{k\to 0} (T)$ (right panel) signal a second-order
phase transition.  Our calculations coincide with chiral perturbation
theory up to $T \sim 45$ MeV.  Around $T_c$ we obtain a scaling
behavior of the order parameter with a critical exponent $\beta
\propto 0.4$ (cf.~right panel).

\section{Critical Behavior}
\label{criticalbeh}

\begin{figure}[htbp!]
\vspace{-0.1in} 
\centerline{\hbox{
\psfig{file=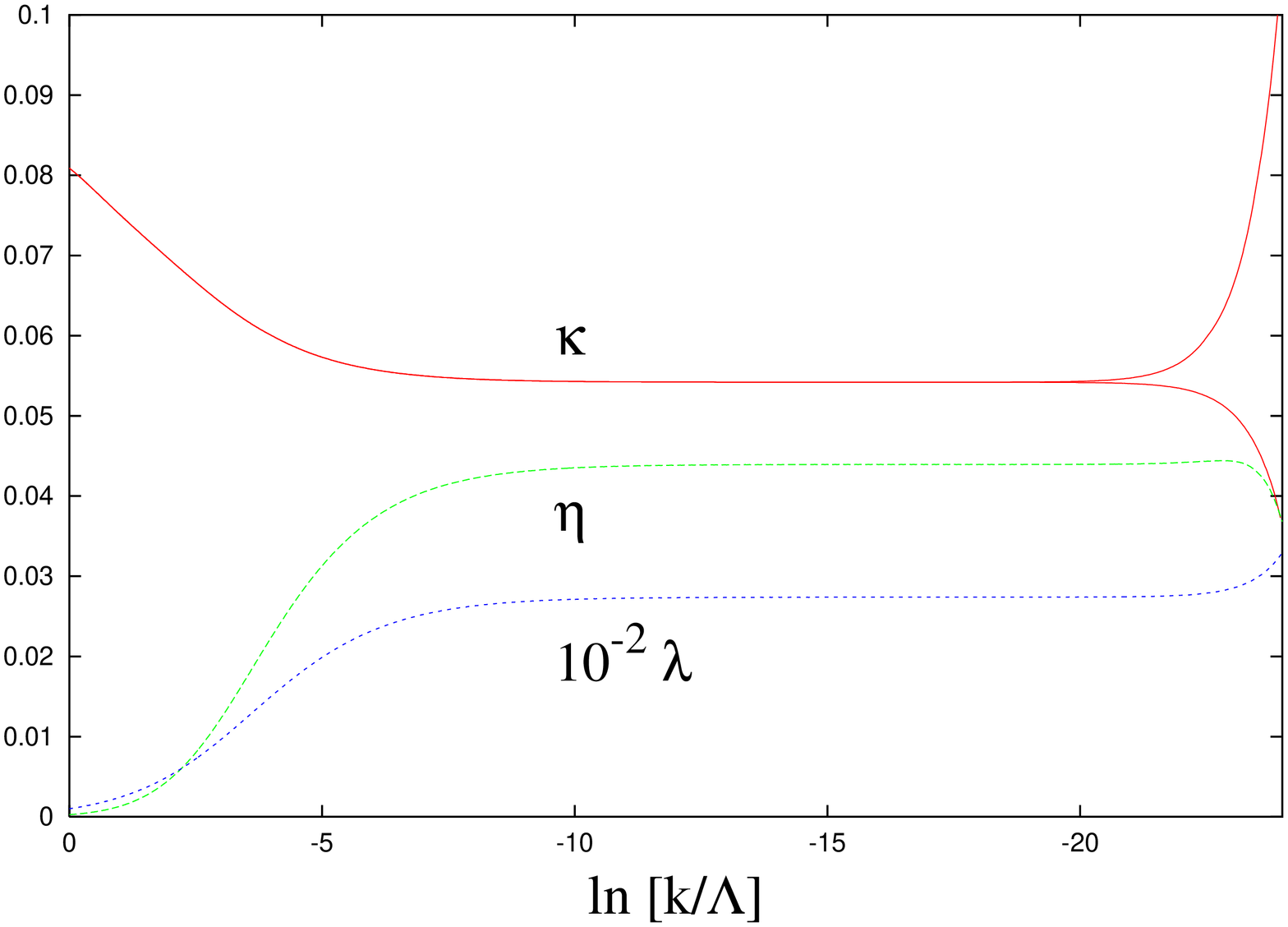,height=1.9in,width=2.4in,angle=-90}
\psfig{file=space.ps,width=0.05cm,angle=-90}
\psfig{file=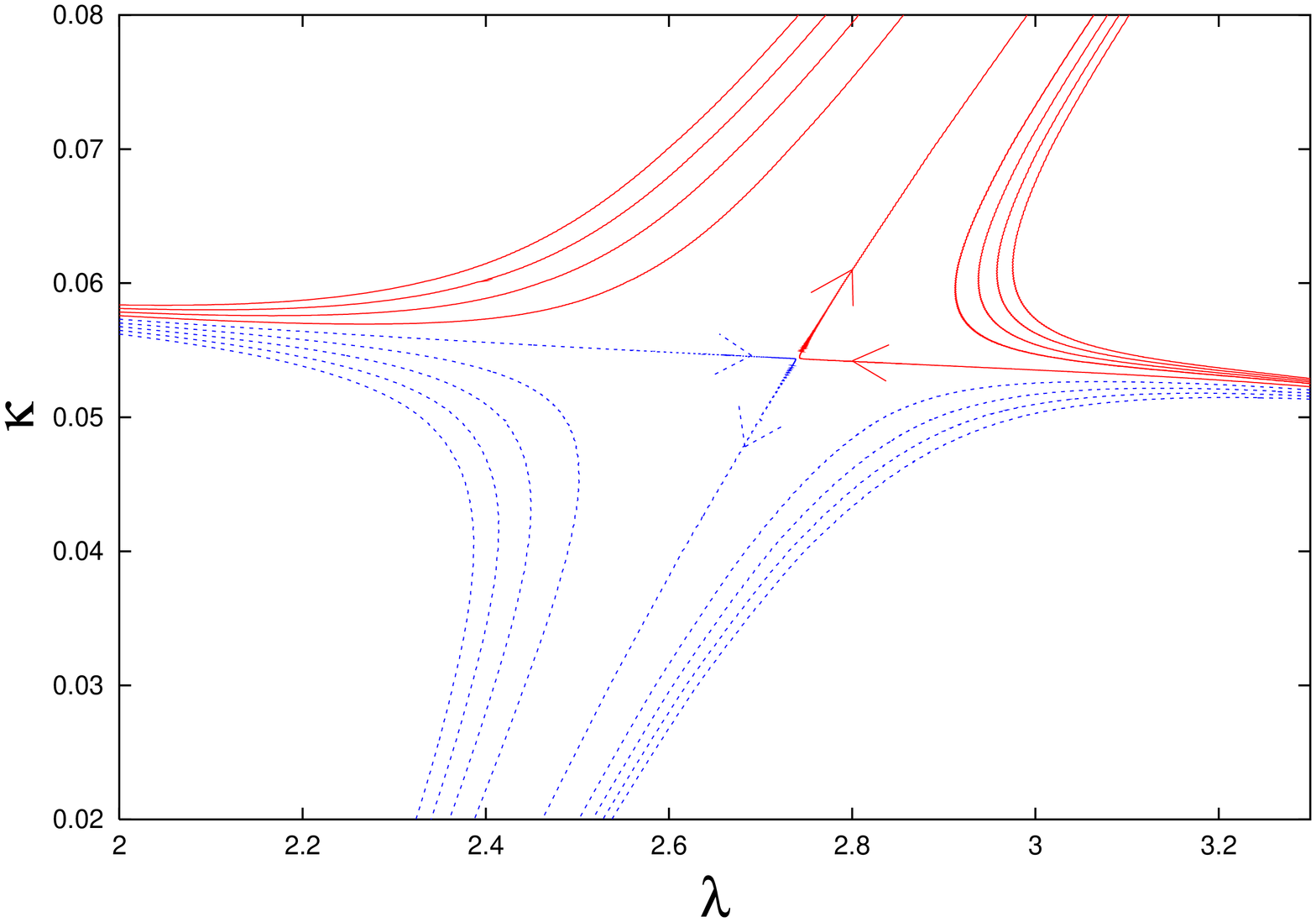,height=1.9in,width=2.4in,angle=-90}
}}
\parbox{12cm}
{\caption{\label{figfixpoint} Left panel: The minimum $\kappa$, $\eta$
and coupling $\lambda$ as function of $t$.  Right panel: The phase
diagram for $N=1$ (see text for details).}}
\end{figure}

At the critical temperature only properly rescaled quantities
asymptotically exhibit scaling and the evolution of the system is
purely three-dimensional~\cite{wett}. This is the well-known
dimensional reduction phenomenon.  The rescaled dimensionless minimum
$\kappa (t)$ of the potential approaches a constant (the fixed point
$\kappa^*$) as the dimensionless flow ``time'' $t = \ln (k/\Lambda) $
tends to $-\infty$ in the infrared (see left panel of
Fig.~\ref{figfixpoint}).  For a starting value near the critical value
$\kappa_{cr}$ at the ultraviolet scale $t=0$ the evolution moves
towards either the spontaneously broken $\kappa \neq 0$ or the
symmetric $\kappa = 0$ phase in the infrared.  Due to the rescaling,
which in the infrared tends to a constant value or zero during the
$k$-evolution, the dimensionless minimum $\kappa = \rho_0/k^2$
diverges for $k \to 0$ in the broken phase.  The time the system
spends on this scaling solution can be rendered arbitrarily long by
appropriate fine tuning of the initial values at $t=0$.

Besides of the trivial high-temperature and low-temperature fixed
points the $O(N)$-symmetric model exhibits a nontrivial mixed fixed
point inherent in the flow equations.  The region around this fixed
point is displayed in right panel of Fig.~\ref{figfixpoint} where the
rescaled dimensionless quantities $\kappa $ and $\lambda $ are plotted
for $N = 1$ and for different initial values during the evolution
toward the infrared.

In general there are two relevant physical parameters for the
$O(N)$-model which must be adjusted to bring the system to the
critical fixed point according to universality class
arguments. Since we here work in the chiral limit (no external sources
or masses) only one relevant eigenvalue from the linearized
renormalization group equations is left (the temperature). This can
also be seen in Fig.~\ref{figfixpoint}. The quantity $\kappa$ is the
relevant variable because repeated renormalization group iterations
(which are the discrete analog to the continuous evolution with
respect to $t \to -\infty$) drive the variable away from the fixed
point value while $\lambda$ is the irrelevant variable being iterated
toward the fixed point if the initial values are chosen sufficiently
close to it.  Thus there is a one-dimensional curve of points
attracted to the fixed point which is the so-called critical surface
for this model.

One nicely sees that this critical line separates both phases. When we
choose initial values near the critical line the system spends a long
time near the critical point. E.g.~starting with $\kappa >
\kappa_{cr}$ and $\lambda$ finite the evolution ultimately tends  to
the zero-temperature fixed point corresponding to very large values of
$\kappa$.  On the other hand starting below the critical line the
evolution of $\kappa$ tends towards the high-temperature fixed point
corresponding to small $\kappa$'s. Only for exactly $\kappa =
\kappa_{cr}$ at $t=0$ the renormalization group trajectory flows into
the critical mixed fixed point, which means that the long distance
behavior at the critical point is the same as that of the fixed
point.

Here the predictive power of this nonperturbative approach becomes
visible: During the evolution near the scaling solution the system
loses memory of the initial starting value in the UV and the effective
three-dimensional dynamics near the transition is completely
determined by the fixed point being independent of the details of the
considered microscopic interaction at short distances. Exactly at a
second-order phase transition the potential should be described by a
$k$-independent solution.  The equation of state drives the potential
away from the critical temperature. As a result, after the potential
has evolved away from the scaling solution, its shape is independent
of the choice of the initial values and exhibits universal behavior.

In the vicinity of the critical point we obtain a scaling behavior of
the system governed by critical exponents which parameterize the
singular behavior of the free energy around the phase transition.
Out of the six critical exponents for the $O(N)$-model only two 
are independent due to four scaling relations which we have verified
by explicit calculations of $\alpha$, $\beta$, $\gamma$ and $\nu$ in
three dimensions for various $N$ (see tab.~\ref{tabcritical}).

The critical exponent $\beta$ parameterizes the behavior of the
spontaneous magnetization or the order parameter for the broken phase
$( \kappa_{\Lambda} > \kappa_{cr} )$ in the vicinity of $T_c$. The
difference $\kappa_{\Lambda} - \kappa_{cr}$ is a measure of the
distance from the phase transition irrespective of any given value of
$\lambda_\Lambda$.  If $\kappa_{\Lambda}$ is interpreted as a function
of temperature this difference is $\propto (T - T_c)$,
i.e.~$\kappa_{cr}$ defines the critical temperature in three
dimensions.

All calculated critical exponents nicely converge to the large--N
values. In addition for $N=1$ we have found $\eta = 0.0439$ and
$\delta = 4.748$.

\begin{table}[ht!]
\begin{center}
\begin{tabular}{|c|c|c|c|c|}
\hline
$ \hspace{4mm}N$ &  $\nu$ & $\beta$ & $\alpha$ & $\gamma$ \\
\hline
N=1   & 0.643  & 0.335  & 0.071  & 1.258 \\
N=2   & 0.695  & 0.3475 & -0.085 & 1.39  \\
N=3   & 0.75   & 0.375  & -0.25  & 1.50 \\
N=4   & 0.79   & 0.395  & -0.37  & 1.58 \\
N=10  & 0.911  & 0.4555 & -0.733 & 1.822 \\
N=100 & 0.993  & 0.4965 & -0.979 & 1.986 \\
\mbox{large--N}& 1.0    & 0.5    & -1.0   & 2.0   \\
\hline
\end{tabular}
\caption{\label{tabcritical} Critical exponents for different $N$ compared with the
large--N result.}
\end{center}
\end{table}
The structure of the regularized RG-improved flow equations depends on
the used $a$ $priori$ unknown cutoff functions $f_k$. Employing the
same class of the cutoff function as in ref.~\cite{papp} we have
calculated critical exponents for several cutoff functions with more
and more monomials on a grid for the full potential.  A small
systematic decrease in the values for the critical exponents with the
number of included monomials of the smearing function is observed.
Restricting to less monomials in the cutoff function $f_k$ seems to
accelerate the IR convergence at the critical point. This is visible
in the evolution of the minimum of the potential which stays longer on
the critical trajectory in the vicinity of the scaling solution.  For
more monomials the evolution of the minimum escapes faster from the
critical trajectory to the symmetric as well as the broken phase.\\

To conclude we have presented a powerful nonperturbative method based
on the RG approach applied to the $O(N)$-model at finite temperature,
which, in fact, is consistent with pertinent lattice simulations. It
remains to be seen in how far the critical $O(4)$-behavior is realized
in full two flavor QCD simulations.
 
\section*{Acknowledgments}
One of the authors (BJS) would like to express his gratitude to the
organizer of the XVII.~Autumn School for financial support and for
providing a most stimulating environment. This work was supported by
NFS-grant NFS-PHY98-00978 and by GSI Darmstadt.

\end{document}